# Digital Library Initiatives in India: A Comprehensive Study

Sankhayan Mukherjee & Swapan Kumar Patra[1]

Department of Library and Information Science

Sidho-Kanho-Birsha University

Purulia, West Bengal PIN- 723149

Email: skpatra@gmail.com; skpatra@skbu.ac.in

**Abstract**

A 'Digital Library' is a specialized library that focuses primarily on the collection, storage, and retrieval of digital objects. The digital collection may be of different format for example, text, audio-video material, photograph and so on. Recently, Government of India has taken several initiatives for digitization of materials. This study is a survey of digital library initiatives in India collecting secondary information from about fifty digital libraries from their respective websites. The findings show that in most cases the actual conception of digital library is still in nascent stage. The online subscriptions and linked to the third-party websites are also considered as a digital library. However, many digital libraries have not any proper search interface in their respective website due to improper arrangement of metadata. In some cases, they do not have their own digitized collection and provided some other's collections or referred to their users to some third-party website. Moreover, there are many digital libraries which cannot be accessed outside (remote access) the organization. Hence, the regular website maintenance, remote access facility and proper training of information professionals are required. Moreover, the so-called digital libraries in India have not developed their own standards or are not following any global standards. However, the usage statistics for the government digital libraries are far better than the usage statistics of academic or public libraries. Users are perhaps more interested in the government's rules, laws, orders etc. That is perhaps a positive sign of digital governance and reaching the public. There are several important observations and policy suggestions that may be helpful for students, scholars, library professionals and the decision makers in the government.

[1] **Corresponding author**

**Introduction**

A specialized library that focuses primarily on the collection, storage, and retrieval of digital objects for example, text, audio-video material, photograph and so on, called 'Digital Library.' A digital library is accessible via the internet or other digital means. The size and scope of a digital library varies depend on its usage and the circumstances where it is stored. It can be created, stored, and maintained by various types of institutions including academic institutions, public libraries, government agencies, and private companies. They may offer free or paid access to their collections, and may provide different levels of access to users based on their needs. The contents stored in a centralized server can be accessed in the stand-alone machine or from the respective institution's local area network, or it can be accessed remotely. Because of their many distinct advantages, digital libraries are becoming increasingly popular. Also, in the 'information society' there is a growing demand for online access to information in every aspect of life.

The concept of Digital Library was first discussed in India in 1996 in a conference held by Society of Information Science, Bangalore (Gurram, 2008)[1]. However, the matter was not much discussed further, at least until about the mid of 2000's. Digital library initiatives were gained popularity in India with the increasing use of ICT and the internet. Recently, Government of India has taken several initiatives for digitization of materials. As a result, the digital library repositories in India have somehow gained a momentum. Among the many, Digital Library of India (DLI) is the biggest national level initiative taken by the Indian government is noteworthy. The DLI is an initiative of the Indian government, and has been developed in partnership with several major libraries and research institutions in India and abroad. It is a part of the Million Book Project or the Universal Digital Library Project, started by Carnegie Mellon University, USA. This notable digital library in India is an online platform that provides access to millions of books and other materials from a variety of sources.

Beside this, the Ministry of Education under its National Mission on Education through Information and Communication Technology has initiated the National Digital Library (NDL) project to develop a framework of virtual repository of learning resources through a single window search facility. It is being developed at Indian Institute of Technology Kharagpur. It is one of the major digital library initiatives in India, which is an online platform that provides access to a wide range of digital resources, including e-books, audio books, videos, images, and other types of multimedia content. The NDLI is an initiative of the Ministry of Education, Government of India, and is designed to promote and support education and research in India.

---

[1]Gurram, S. (2008). Digital library initiatives in India: a proposal for open distance learning.

Beside this, Digital Library of Art Masterpiece, National Mission for Manuscripts, Vidyanidhi, National Institute of Advanced Studies (NIAS), Bangalore, etc. are also noteworthy digital library initiatives in India. With all these efforts India has a rapidly growing digital library landscape, with both government and private organizations investing in the development of digital libraries.

In addition to these national initiatives, there are also many academic and research institutions in India that have developed their own digital libraries, such as the Indian Institute of Technology (IIT) Digital Library and the Jawaharlal Nehru University (JNU) Digital Library and so on.

Overall, the digital library landscape in India is rapidly expanding, with more and more organizations investing in the development of digital resources and platforms to support education, research, and access to information.

The digital library collection and maintenance is not an easy task. Its value depends on the users and their usage for that library. If the users get satisfactory services from the library, then the library is perhaps useful. However, in Indian scenario, there are many serious issues in the digital libraries regarding the storage, access of content and services. In this context, this study is a brief survey of various digital library initiatives taken all over India.

The study aims to survey the major digital library initiatives in India from their respective website. The study is going to observe the digital library initiatives by using the following parameters: geographical locations, digital collections, software used for retrieval by that institution, usage statistics, copyright issues, remote access facility, social media presence etc. From above mentioned parameters, this study will examine the present status and the progress of digital library initiative allover India.

The study is original in the sense that this is a comprehensive survey covering all major digital libraries in India. During this study, we observed that articles in this area mainly focused on the digitization process and digitization effort of single institutions. Nevertheless, some of them discussed about the digitization problems and copyright issues. However, expect a few (Das 2022) there was no significant study about the digital library initiative in all over India. So, this is an effort to fill the gap in understanding the digital library landscape of India.

**Literature Review**

Physical documents can be preserved for a much longer period through digital preservation. Hence the use of digital library will be increased at a proportional rate (Jeevan, 2004)[2]. In developed countries, for example in the US, in the mid 2000's there was already 60%-70% of data available in digital format. At the same time, in developing countries, like India, there was only 2.5%-3% data available in digital format (Kaur & Singh, 2005)[3]. Nevertheless the 'Digital Library' concept in Indian libraries is still being discussed about internal operations (using databases on electronic media such as CD-ROMs), access to the subscribed web journals, free resources etc. (Jeevan, 2004). All the project on Digital Library Initiative has been done so far in India only focused on creating specialized digital collections, such as art, culture, and heritage of India (Jain & Babbar)[4]. However, the digital library initiatives in India were in the germination stage at that time.

In the case of developing countries, the main problem of digital libraries is in identifying valuable contents for digitization and lack of proper financial support and infrastructure from the respective institutions. This was because of lack of goodwill, technical infrastructure for digital library, copyright policy and adequate funds for building a proper digital library. Discussing about the issues of Digital Libraries in India, Sreekumar & Sreejaya (2005)[5] focused on the problems of building a digital library as lack of information & communication technology (ICT) infrastructure, proper planning and integration of Information resources, ICT strategies and policies, technical skills, rigidity in the publishers' policies and data formats, management support and copyright/ IPR issues. The most surprising thing is that, in North-East India, some of the libraries expressed themselves as digital libraries where they had only web-OPAC system (Mukherjee & Patra 2022)[6]. From this, it can be understood that the knowledge about digital library is still unclear in India. Nevertheless Bhattacharya (2004)[7] attributed the digital divide facing by India as a problem in case of developing digital library.

---

[2]Jeevan, V. K. J. (2004). Digital library development: identifying sources of content for developing countries with special reference to India. *The International Information & Library Review*, *36*(3), 185-197.
[3]Kaur, P., & Singh, S. (2005). Transformation of Traditional Libraries into Digital Libraries: A Study in Indian Context. *Herald of Library Science*, *44*(1/2), 33.
[4]Jain, P. K., &Babbar, P. (2006). Digital libraries initiatives in India. The International Information & Library Review, 38(3), 161-169.
[5]Sreekumar, M. G., &Sreejaya, P. (2005). Digital Library Initiatives and Issues in India: efforts on scholarly knowledge management.
[6]Mukherjee, S., & Patra, S. K. (2022). Digital Library Initiatives in North East India: A Survey. *Indian Journal of Information, Library and Society*, 35(3-4), 203-218.

The digital library evolution is not an easy task. Its value depends on the users and their usage for the library (Xie, 2008)[8]. So, the relationship between the digital library uses and the digital library evolution is totally depending on the user satisfaction in terms of their expectations and desired image for digital library. As Saracevic (2000)[9] discussed about the challenges facing by digital library evolution process and provided a conceptual framework for overcoming the actual problems. To solve the fundamental problem of digital library evolution, the first thing to do is to find the history of digital library (Castelli & Pagano[10] 2012). The growth of Library and Information Science literature in India and the facilities for LIS education and research (Patra & Chand 2006) [11]it can be assumed that the digital library could have been a difficult task. So, the way to give momentum to the initiation to the digital library development in India, requires two-pronged strategy, including both the digitization of local content in one hand and access to external resources on the other hand.

It is a very promising thing that, in present days Indian Government and some of the state governments have taken the initiatives for development and progression of digital library in India. As Bhattacharya (2004) explained the earlier initiatives taken by the Indian Government, such as the INDEST Consortia and so on. The Ministry of Human Resource Development (MHRD) has launched a "Consortia-based Subscription to Electronic Resources for Technical Education System in India" named as the Indian National Digital Library in Science and Technology (INDEST) Consortium (Arora & Agrawal, 2003)[12].Deb (2006)[13] discussed about The Energy and Resources Institute (TERI), a non-government initiative. Researchers get single window access of better structured data from their desktop in this institute.

---

[7]Bhattacharya, P. (2004). Advances in digital library initiatives: a developing country perspective. The International Information & Library Review, 36(3), 165-175.
[8]Xie, H. I. (2008). Users' evaluation of digital libraries (DLs): Their uses, their criteria, and their assessment. Information Processing &Amp; Management, 44(3), 1346–1373.
[9]Saracevic, T. (2000). Digital library evaluation: Toward an evolution of concepts
[10]Candela, L., Castelli, D., & Pagano, P. (2012). History, evolution, and impact of digital libraries. In Organizational Learning and Knowledge: Concepts, Methodologies, Tools and Applications (pp. 837-866). IGI Global.
[11]Patra, S. K., & Chand, P. (2006). Library and information science research in India: A bibliometric study.
[12]Arora, J., & Agrawal, P. (2003). Indian Digital Library in Engineering Science and Technology (INDEST) Consortium: consortia-based subscription to electronic resources for technical education system in India: a Government of India initiative.
[13]Deb, S. (2006). TERI integrated digital library initiative. The Electronic Library, 24(3), 366-379.

In this research area, several research papers discuss with the institutional level digital library initiatives (Das 2022)[14]. Such as Gaur (2003)[15] consulted with the library automation state in Indian Management Institutes in India, Mujoo-Munshi (2003)[16] discussed about the building procedure of digital resources in the Indian National Science Academy (INSA), Joshi (2006)[17] has discussed about the Tocklai Central Library digitization initiative of the institute etc. These types researches can be considered as case study of digitization from single institute. From all these studies, we mainly observed that there is a huge problem about the resource supply to fulfill the user's needs. This problem emerges more clearly from the study of Naga, Pradhan, Arora & Chand (2008)[18], as they have discussed about the usage trends of e-journals in the universities of North-East India. The usage statistics and status of resource supply status have been dealt in a study of Mukherjee & Patra (2022) also. In the context, this study is going to fill avoid of the study on the digital library initiative in India from the data derived from the different websites of digital library initiatives in India. While doing so, this paper has the following research objectives.

**Research Objectives**

The study is an exploratory study trying to investigate the contemporary scenario of digital library initiatives in India. The study is going to investigate the present status of digitization in the digital libraries in India in terms of their location, numbers of collections, software used for digitization, access mode, search interface, social media presence and remote access facility and so on.

**Methodology**

For a comprehensive survey of digital library initiative in India, secondary information of about 50 digital libraries in India are collected from their respective websites. A database has been created from the collected data using the MSExcel software. The sample includes major digital libraries of various nature including academic, government and other types of libraries.

---

[14] Das, A.K. (2022), "A brief overview of recently launched digital libraries of India", Library Hi Tech News, Vol. 39 No. 2, pp. 18-20. https://doi.org/10.1108/LHTN-11-2021-0085

[15] Gaur, R. C. (2003). Rethinking the Indian digital divide: the present state of digitization in Indian management libraries. *The International Information & Library Review*, *35*(2-4), 189-203.

[16] Mujoo-Munshi, U. (2003). Building digital resources: creating facilities at INSA. *The International Information & Library Review*, *35*(2-4), 281-309.

[17] Joshi, G. (2006). Digital Library Initiative in North East India with Special Reference to Tocklai Experimental Station: A Case Study.

[18] Naga, M. M., Pradhan, D. R., Arora, J., & Chand, P. (2008). Access to e-journals through UGC Infonet Digital Library Consortium: A study of usage Trends among the Universities of North East India.

**Results**

The selected website of the Indian digital libraries was accessed in the month of November 2022 to February 2023. During those four months the websites were accessed in every week. The data were collected, analyzed, and presented in the following section.

**Sample description**

A total of fifty digital libraries from different zones of India is considered for this study. In the sample, there are eight digital libraries from the east zone, eleven from west zone, thirteen from south zone, sixteen from north zone and three from central zone. The libraries are also differentiated in the following four types, Academic, Government, Public and autonomous bodies. There are twenty-four academic, two autonomous, twelve government and twelve public libraries are considered in this study.

*Year of establishment*

From the database, the year of most of the digital libraries was not available. The earliest among the sample is Digital library of Developing Library Network (DELNET), established in the year 2000. The Tamil Nadu Dr. M.G.R. Medical University (Tamil Nadu) was established in 2002. Besides these, Inclusive Digital Library, KarnaVidya Foundation (Tamil Nadu); Panjab Digital Library (Chandigarh); National Mission for Manuscripts (Delhi) and Muktabodha Indological Research Institute (MIRI) (Delhi) were established in the year of 2003. Moreover, Digital Library of the West Bengal Secretariat, Government of West Bengal and National Digital Library of India were established in the year of 2011 and 2018 respectively.

*Geographical locations*

From the database, northern region of India has more numbers of digital libraries with respect to other region of India (figure 1). There are total of 18 digital libraries in northern region of India. There are 3 digital libraries in central region (all are in Madhya Pradesh, a central province in India), 8 in the eastern region (one each in Assam and Bihar, two in Orissa and four in West Bengal), 18 in the northern region (one each in Chandigarh, Haryana, Himachal Pradesh, and Jammu& Kashmir, eight in Delhi, two in Uttarakhand and four in Uttar Pradesh), 13 in Southern region (7 in Tamil Nadu, 4 in Kerala, 2 in Karnataka) and 9 in west region (5 in Gujarat and 4 in Maharashtra).

**Figure: 1 Location of major digital libraries in India**

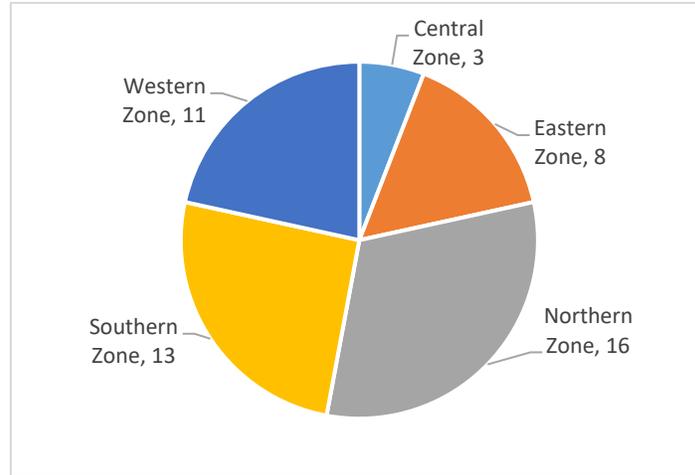

*Types of digital Libraries*

Based on their hosting organization, digital libraries can be categorized into four types. In the first category there are libraries hosted by the academic institutions, public libraries with general documents and the libraries with government documents. In the database there are 24 libraries that are hosted by academic institutions containing academic documents, 15 libraries contain general documents and 9 libraries contain government documents. Overall, these collections contain digitized books, theses, CD, public documents, manuscripts, journals, newsletter and so on.

**Figure: 2 Types of Digital Libraries based on their collections**

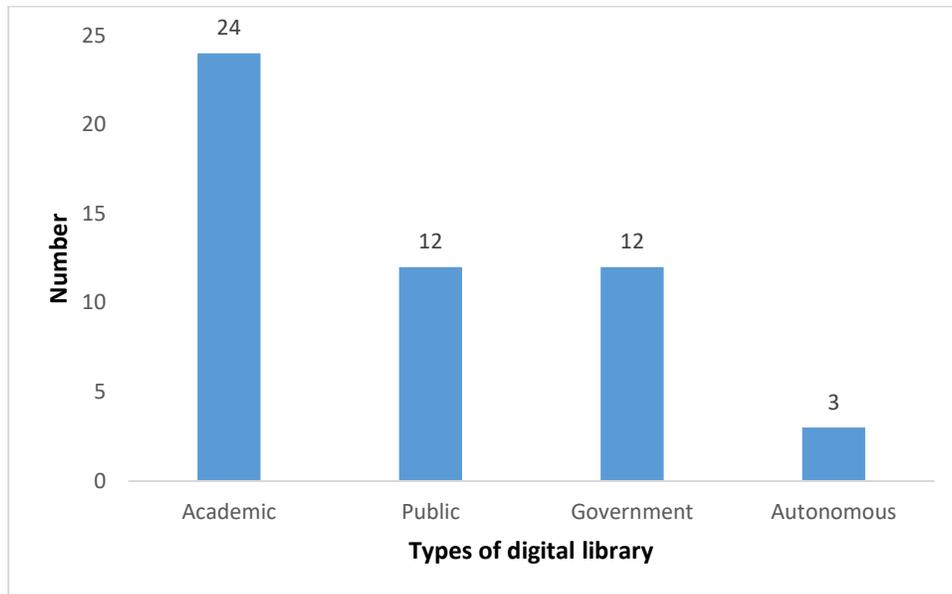

*Software used by the digital library*

There is not much information available about the software used for either digitization, storage, or metadata creation. There are about 11 libraries (out of 51 from the database) displayed information about the software, the libraries are using for the digitization, in their websites. Most of the libraries are using DSpace, an open-source software for the storage and metadata creation. The libraries which use DSpace are Raj Bhavan, Assam; Biju Patnaik Central Library; C H Mohammed Koya Library, University of Calicut; Library and Information Centre, Indian Institute of Management Kozhikode; Directorate of Public Libraries – School Education Department, Government of Tamil Nadu; Smt. Hansa Mehta Library, The Maharaja Sayajirao University of Baroda; Digital Library, K.E. Society's Rajarambapu Institute of Technology, Rajramnagar; Learning Resource Center, Indian Institute of Technology, Indore; Digital Library, Lok Sabha, Parliament of India and Archives of Indian Labour, V.V. Giri National Labour Institute. There are a few libraries use both DSpace and Greenstone. Those libraries are Sir Dorabji Tata Memorial Library, Tata Institute of Social Sciences.

*Access Mode*

The digital libraries use three types of access mode to access their collections. Most of the libraries use open access mode, followed by close access and semi open access. There are 30 libraries (59%), out of 51, allow open access, 16 libraries (31%) have close access system, 4 are close access system (about 8%) and 1 library (2%) is holding both close and open access mode. Here, we mentioned semi open access system as, if someone want to access content from those websites, then he must have to create account in those sites. Now, the libraries, who are holding close access system, may have paid version or they only give access their contents to their own user (such as some educational institution give the access of their contents to their students only). In open access system, anybody can access the contents of the particular library without any resistance and condition. (Figure 3)

**Figure: 3 Access mode**

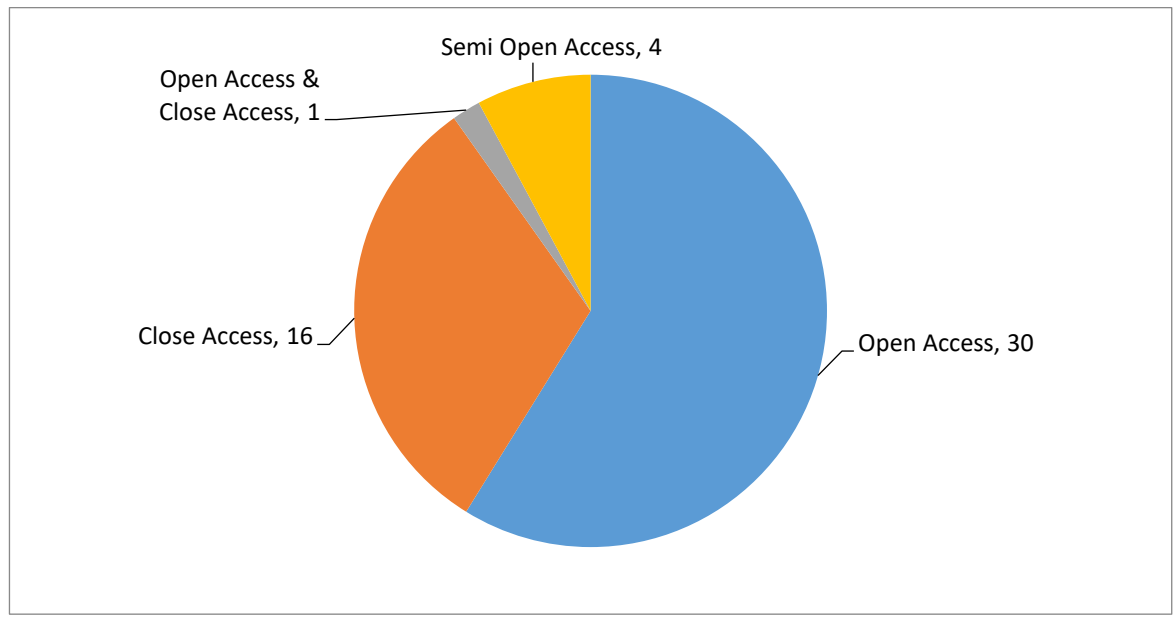

*Remote Access Facility*

In this category, we define remote access facility as, how we can access the library website during our study. When we could access the library website and accessed their contents easily, then we considered that, the library has remote access facility; and when we did not access their contents from our retrieving system, then we considered that, the library is not holding remote access facility. During our study, we differentiated 51 libraries in three categories, as who had remote access facility, who had no remote access facility and who used some third-party website to provide their contents to their user. During our study, we found 41 libraries, who had remote access facility, 5 libraries did not hold remote access facility and 5 libraries used some third-party website. (Figure 4)

**Figure: 4 Remote access facility**

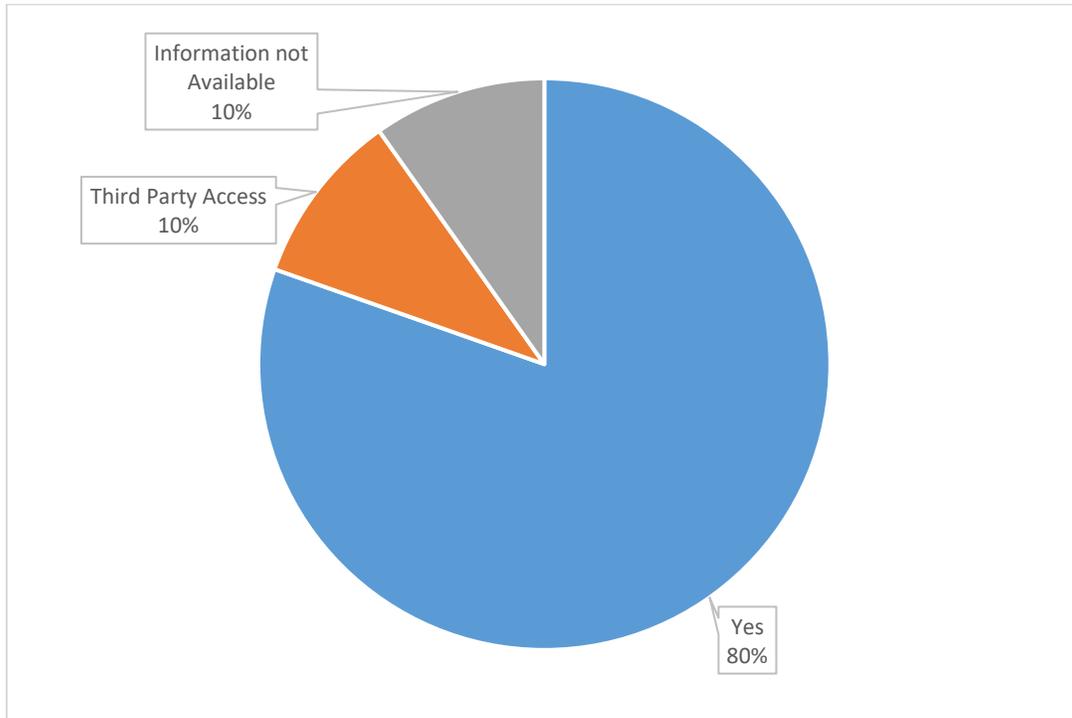

*Copyright Policy*

A digital library system should have some copyright policy. From our constructed database about the digital library initiative in India, we found that, majority of libraries had no stated copyright policy for their digital collection. Though, some of them had some 'volatile type' of copyright policy and a small number of libraries had some proper copyright policy. In this study sample only 9 digital libraries out of 51, has stated copyright policy. Rest of the 42 libraries had no copyright policy for their collections.

*Search Interface*

Search Interface is main criteria for a digital library and its uses. A user-friendly search interface can upgrade a digital library in the next level. From the database, it is observed that 35 libraries out of 51 had search interface in its website, but rest 16 libraries had not preferable search interface in their websites (Figure: 5)

**Figure: 5 Search interface**

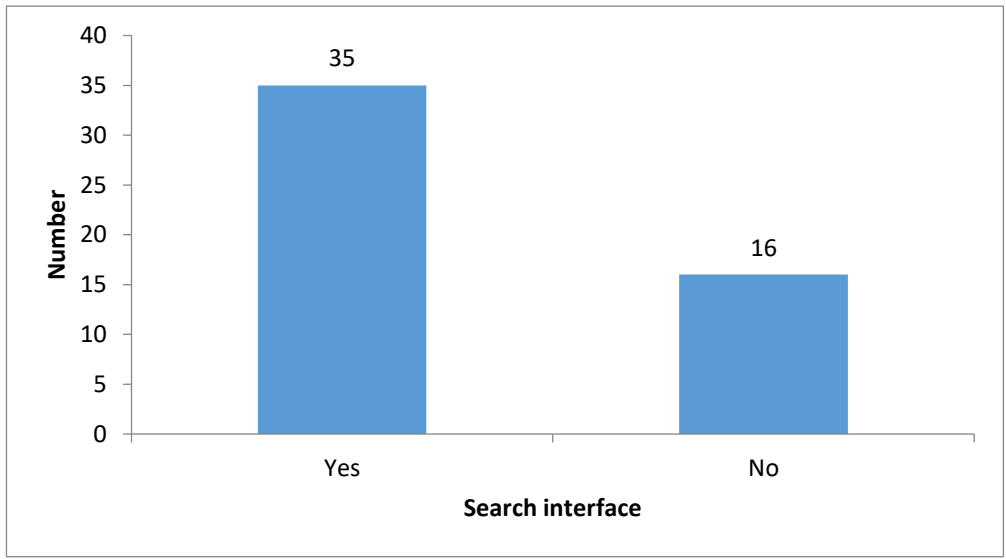

*Social Media Presence*

Today, social media has all pervasive impact in the human life. Libraries all over the world are using various social media platform to do marketing of their product and services. Therefore, if a digital library can make their presence to the public through social media platform, it could reach people quite easily. In this sample, a total of 24 library out of 51 have their social media presence in prominent social media platforms (for example, Facebook, Twitter, YouTube etc.). The rest 27 (53%) do not have any social media presence (Figure 6).

**Figure: 6 Social media presence**

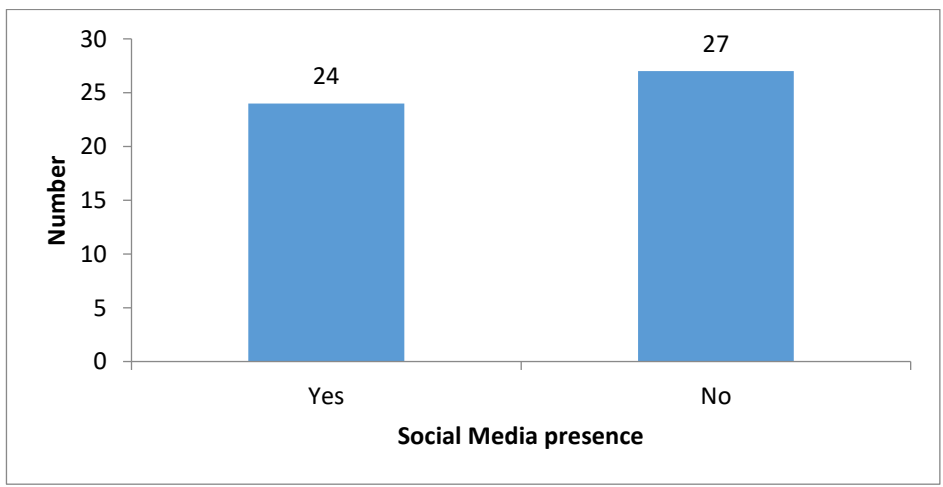

**Discussion**

The findings show that in India, most of the cases the actual conception of digital library is still in nascent stage. The online subscriptions and linked to the third-party websites are also considered as a digital library. As per the definition, a digital library is developed by converting hard copy materials, through scanning, to digital objects and accessed from computers. However, many of these types of libraries can be considerd as "E-library" which are the collections of offline and links to online resources and many of them are 'born-digital'.

Moreover, many digital libraries have not solid search interface in their respective website due to improper arrangement of metadata. In some cases, they have not even their own digitized collection and provide some other's collections or referred to their users to some third-party website. There are many digital libraries which cannot be accessed outside (remote access) the organization because of various reasons. Hence, the lack of regular website maintenance, remote access facility and proper training of library professionals are required. Moreover, the so-called digital libraries in India have not developed their own standards or are not following any global standards. In contrast, the digital libraries of developed countries, are complying with some specific well established digital library standards. During our study, we observed that the usage statistics for the government libraries are far better than the usage statistics of academic and public libraries. So, we can conclude that people are more interested in the government's rules, laws. orders etc. That is perhaps a positive sign of digital governance.

**Conclusion**

Digital libraries can offer several benefits compared to traditional libraries. For example, they can provide users with 24/7 access to materials from all over the globe. These types of libraries are different from the traditional libraries in terms of faster and more efficient searches and information retrieval. Additionally, digital libraries can store huge amounts of information in a relatively small space, making it easier to store and manage large collections.

However, there are some challenges associated with digital libraries, such as the need for ongoing maintenance and updating. It is an enormous challenge to ensure the security and privacy of users' data. It also needs regular updating, effective information management and organization. Despite these challenges, digital libraries continue to grow and evolve, providing users with an increasingly diverse and comprehensive range of digital resources.

At per with the global initiatives, the digital library initiative began in India in the mid-1990s. However, it got the momentum almost after a decade. After IT and the internet came to Indian market, the actual digitization process speeded up. Although, being a developing country, Indian initiatives are praiseworthy, it is lagging the developed countries, particularly the Europe, where the concept of digitization was established long ago.

The digital library initiative started in India in the mid-1990s, the initiative has not made a strong foundation till date. Many organizations, that are working on library networking are taking initiatives for development of digital libraries in India. Amongst all other agencies, INFLIBNET has been playing a significant role for the modernization of university and college libraries of India.

However, it appears that users are facing problems in most of the cases while accessing the digital contents. The main issue here is the absence of proper management system, due to the lack of good knowledge about the maintenance of the digital library. Many of those libraries only subscribe e-journals and provide users to get online access from the subscribed websites through the institution's website to get the e-contents. It was also observed that, many of the websites of the digital libraries were not working due to broken links, website crash etc. This is perhaps due to the lack of knowledge and professionalism of those libraries in both the planning and implementations of the digital library.

Efficient search interface and copyright policy are the two main things for building a digital library. This study has observed that, many of the libraries did not have robust search interface. Also, they do not follow or do not mention any copyright policy or related issues.

Many libraries have the closed access for their digital collection, where as many of the libraries provide open access digital content to its users. Although, there are some developments in terms of digitization, it is perhaps not a premature claim that India has not progressed much in developing a digital library in a true sense. Hence, the study recommends that a national policy should be adopted to keep a uniformity to benefit all stakeholders.

This study is based on the selected digital libraries in India. This is also based on the secondary information sources, particularly the websites on the respective libraries. A comparative study of digital libraries including libraries from other developing or developed parts of the world would probably give a further holistic picture of digital library initiatives in India.